\pdfoutput=1
\documentclass[a4paper,11pt]{article}
\usepackage{amsfonts}
\usepackage{mathrsfs}
\usepackage{amsmath}
\usepackage{amssymb}
\usepackage{framed}
\usepackage[medium]{titlesec}
\usepackage{bm}
\usepackage{cite}
\usepackage[normalem]{ulem}
\usepackage{slashed}
\usepackage{isodateo}
\usepackage{graphicx}
\usepackage{xcolor}
\usepackage[bookmarksnumbered=true,bookmarksopen=true]{hyperref}
\usepackage{pbox}

 \hypersetup{colorlinks,%
             linkcolor=[rgb]{0,0.3,0.6}, %
             citecolor=[rgb]{0,0.3,0.6}, %
             urlcolor=[rgb]{0,0.3,0.6}}
\usepackage[hmargin=.7in,vmargin=1.1in]{geometry}
\usepackage{indentfirst}
\renewcommand{\rm}{\mathrm}

\graphicspath{{fig/}}

\def\bge{\begin{equation}}
\def\ede{\end{equation}}
\def\bga{\begin{aligned}}
\def\eda{\end{aligned}}
\def\bgp{\begin{pmatrix}}
\def\edp{\end{pmatrix}}
\def\bgs{\begin{subequations}}
\def\eds{\end{subequations}}

\def\di{{\mathrm{d}}}

\def\Ord{{\mathcal{O}}}

\def\to{\rightarrow}

\def\sec{\mathrm{s}}
\def\cm{\mathrm{cm}}

\def\keV{\mathrm{keV}}

\def\di{\mathrm{d}}

\def\dm{\mathrm{dm}}
\def\crit{\mathrm{crit}}
\def\eq{\mathrm{eq}}

\def\fp{f_\phi}
\def\np{n_\phi}
\def\mp{m_\phi}
\def\sigv{\langle \sigma v \rangle}

\def\bge{\begin{equation}}
\def\ede{\end{equation}}
\def\bga{\begin{aligned}}
\def\eda{\end{aligned}}
\def\bgp{\begin{pmatrix}}
\def\edp{\end{pmatrix}}
\def\bgs{\begin{subequations}}
\def\eds{\end{subequations}}

\usepackage{enumerate}

\makeatletter
\newcommand{\xRightarrow}[2][]{\ext@arrow 0359\Rightarrowfill@{#1}{#2}}
\makeatother

\newcommand{\beq}{\begin{eqnarray}}
\newcommand{\eeq}{\end{eqnarray}}
\newcommand{\bea}{\begin{eqnarray}}
\newcommand{\eea}{\end{eqnarray}}

\newcommand{\nn}{\nonumber}
\def\OMIT#1{{}}
\newcommand{\lsim}{\mathrel{\rlap{\lower4pt\hbox{\hskip1pt$\sim$}}
    \raise1pt\hbox{$<$}}}         
\newcommand{\gsim}{\mathrel{\rlap{\lower4pt\hbox{\hskip1pt$\sim$}}
    \raise1pt\hbox{$>$}}}         

\begin{document}

\title{\Large\textbf{Galactic Origin of Relativistic Bosons and XENON1T Excess}}
\author{Jatan Buch\footnote{Email:\texttt{ jatan\char`_buch@brown.edu}}, ~Manuel A. Buen-Abad\footnote{Email:\texttt{ manuel\char`_buen-abad@brown.edu }}, ~JiJi Fan\footnote{Email:\texttt{ jiji\char`_fan@brown.edu}}, ~and~John Shing Chau Leung\footnote{Email:\texttt{ shing\char`_chau\char`_leung@brown.edu}}\\[2mm]
\normalsize{\emph{Department of Physics, Brown University, Providence, RI 02912}}}

\date{}

\maketitle

\begin{abstract}
We entertain the exotic possibility that dark matter (DM) decays or annihilations taking place in our galaxy may produce a flux of relativistic very weakly-coupled bosons, axions or dark photons. We show that there exist several upper bounds for this flux on Earth assuming generic minimal requirements for DM, such as a lifetime longer than the age of the Universe or an annihilation rate that leaves unaffected the background evolution during matter domination. These bounds do not depend on the identity or the couplings of the bosons. We then show that this new flux cannot be large enough to explain the recent XENON1T excess, while assuming that the bosons' couplings to the Standard Model are consistent with all current experimental and observational constraints. We also discuss a possible caveat to these bounds and a route to explain the excess.

\end{abstract}

\section{Introduction}
The XENON collaboration recently reported results from searches for new physics using low-energy electronic recoil data with an exposure of 0.65 ton-years~\cite{Aprile:2020tmw}. They observe an excess of $53 \pm 15$ events (a 3.5$\sigma$ Poisson significance) over the known background in the energy bins between (1--7) keV with a peak between (2--3) keV. An intriguing explanation for the excess, proposed by ref.~\cite{Aprile:2020tmw}, is the solar axion model in which relativistic axions from the Sun with energy in the keV range are absorbed by the detector. Although experimental anomalies come and go, they motivate us to think about new theoretical ideas and experimental opportunities that might have been previously overlooked.

Broadly, there could be at least four possible routes to explain the XENON1T excess using new physics: {\it a)} absorption of relativistic bosons with keV scale energy such as solar axions (which has been a benchmark scenario for direct detection experiments~\cite{Arisaka:2012pb, Aprile:2014eoa, Akerib:2017uem, Fu:2017lfc}); {\it b)} scattering of relativistic particles with keV scale energy such as solar neutrinos with an enhanced magnetic moment~\cite{Bell:2006wi, Aprile:2020tmw} or with new interactions~\cite{Boehm:2020ltd}; {\it c)} absorption of a non-relativistic dark matter (DM) particle with mass about (2--3) keV~\cite{Aprile:2020tmw, Takahashi:2020bpq, Alonso-Alvarez:2020cdv}; {\it d)} scattering of non-relativistic particles off electrons, either thermally produced dark matter particles scattering through a semi-annihilation-like process \cite{Smirnov:2020zwf} or boosted dark matter particles with a speed $\sim25$ times the escape speed of our galaxy \cite{Kannike:2020agf, Fornal:2020npv}. The XENON preprint suggests that the fit of a peak-like excess as predicted in scenario {\it c)} above is statistically less significant.\footnote{The preprint does not quote an explicit number, but only states that the global significance of scenario {\it c)} is less than $3\sigma$. This may mainly be due to the fact that scenario {\it c)} is subject to a look-elsewhere effect while explanations of solar axion/neutrinos do not suffer from this effect since their energies are determined by the solar temperature, which is $\sim\Ord(\keV)$.}

In this article, we will focus on scenario {\it a)}: absorption of relativistic bosons, $\psi$, either axions or dark photons, leading to an ionization signal in a direct detection experiment. One serious issue facing such a scenario is that given the current stringent constraints on $\psi$'s couplings to standard model (SM) particles, it is hard to obtain a large enough rate to explain the excess. Indeed, as already shown in the XENON collaboration report~\cite{Aprile:2020tmw}, for solar axions (probably the most appealing scenario since its energy range is pinned by the SM physics determining the solar temperature), the axion-SM couplings required to explain the excess are in conflict with various constraints (e.g., from the red giant branch on the axion-electron coupling~\cite{Giannotti:2017hny}) by at least one order of magnitude, irrespective of the combination of the various couplings considered. In other words, if we assume the axion-electron coupling saturates the red giant bound, the predicted number of events from absorption of ABC solar axions\footnote{ABC stands for the main sources of axion production in the Sun: Atomic recombination and deexcitation, {\it Bremsstrahlung}, and Compton.} will be reduced by a factor of $10^{-4}$, {\it i.e:} we would only predict about 0.01 events.

One potential way to evade the existing observational bounds is to decouple the production of $\psi$ from its absorption process in direct detection. More concretely, $\psi$ may dominantly come from a source different from the Sun, which relies on coupling of $\psi$ to the SM particles. For example, novel ways of producing $\psi$ are DM decays or annihilations. In this case, the production of $\psi$ relies on its couplings to a dark sector, which may leave more room for a possible explanation of the XENON1T reported excess. In particular, one could imagine a scenario in which the dark sector does not couple to any SM particles directly except through a $\psi$ portal, potentially causing the couplings of the dark sector to $\psi$ to be weakly bounded.

In this article we devote Sec.~\ref{sec_bounds} to show that in generic models there are several upper bounds for the flux on Earth of relativistic bosons $\psi$, whether the source for this flux is DM decays or DM annihilations. We then use these conservative bounds to demonstrate in Sec.~\ref{sec_xenon1T} that even if there were such a source, it is impossible to explain the XENON1T excess for $\psi$-SM couplings that saturate all current constraints. We discuss a possible workaround to the bound in Sec.~\ref{sec_caveats}, which requires more convoluted model building and could still be subject to both model-dependent and model-independent constraints. Finally, we present our conclusions in Sec.~\ref{sec_disc}.

We want to emphasize that while our work is motivated by the XENON1T excess, the bounds we derive on the generic scenario of relativistic bosons from DM decays or annihilations are independent of the excess.

\section{Upper Bounds on the Flux of Relativistic Particles from DM Decays or Annihilations}
\label{sec_bounds}

In this section we consider the flux on Earth of relativistic particles originating from either DM decays or annihilations. We will assume that the mean free path of these particles is longer than the distance between the source of this flux and Earth, an assumption realized in the axion and dark photon cases. We will then show that, based on several generic requirements of DM, there are upper bounds on this flux. In this section we present our results in terms of an arbitrary DM mass, but in the subsequent sections we will focus on scenarios with energy scales around keV.

\subsection{DM Decays}
\label{sec_decays}

First we consider the scenario of the DM particle $\phi$ with mass $m_\phi$, decaying to relativistic particles, $\psi$, with a lifetime $\tau_\phi$. The flux at Earth is then given by
\beq
\frac{d\Phi_d}{d E} = \left(\frac{dN_\psi}{dE}\right)_0 \frac{f_\phi}{4\pi \tau_\phi m_\phi} \int \rho_{\rm{DM}}(s) \di s \di \Omega,
\label{eq:decay0}
\eeq
where $\left(\frac{dN_\psi}{dE}\right)_0$ is the energy spectrum of the produced $\psi$; $f_\phi$ is the fraction of dark matter being $\phi$ and $s$  is the heliocentric distance. Note that the integral above is the usual $J$-factor for DM decays, $J_{\rm{dec}} =  \int \rho_{\rm{DM}}(s) \di s \di\Omega$, widely used in indirect detection. Note that in our case we integrate over the entire sky, while in indirect detection the signal usually comes from a specific region of interest, often only a part of the sky. It turns out that the dominant contribution to the $J$-factor is from the Milky Way (MW) halo, which is about $10^{23}$ GeV cm$^{-2}$. More details of calculations on the $J$-factor can be found in Appendix~\ref{app_Jfactor}.

The lifetime of the decaying DM has to be at least longer than the age of the Universe, $\tau_\phi \lesssim 4 \times 10^{17}$ s. With this simplest model-independent requirement, the maximal flux on Earth of the relativistic particles from DM decays is
\beq
\Phi_{d} \approx 10^{10} \, {\rm{cm}}^{-2}{\rm{s}}^{-1} f_\phi \left( \frac{4~\keV}{m_\phi} \right) \left( \frac{4 \times 10^{17} \, {\rm{s}}}{\tau_\phi}\right) \left[\int \left(\frac{dN_\psi}{dE}\right)_0 dE\right].
\label{eq:decay}
\eeq
For simple two-body decays with both daughter particles being $\psi$, $\int \left(\frac{dN_\psi}{dE}\right)_0 dE=2$. More generally, this integral yields an ${\cal{O}}(1)$ number. If DM decays to standard model particles, the constraints on its lifetime usually are significantly stronger. Depending on the final states, the DM lifetime may need to be much longer than the age of the Universe. There are also strong constraints on DM decaying to only relativistic particles from the CMB and large scale structure observations, independent of the final states~\cite{Audren:2014bca, Blackadder:2014wpa, Enqvist:2015ara, Blackadder:2015uta, Poulin:2016nat, Clark:2020miy}. One could also consider a more complicated scenario where each DM particle decays into a non-relativistic daughter and a relativistic one. The minimum DM lifetime in such a scenario still turns out to be slightly longer than the age of the Universe~\cite{Abellan:2020pmw}. Note that in Eq.~\eqref{eq:decay}, we include the parametric dependence of the flux on the decaying dark matter fraction. Given the constraint derived from Planck data in ref.~\cite{Poulin:2016nat}, $f_\phi/\tau_\phi < 2\times 10^{-19}~\sec$, the maximum flux is about $10^{9} \, {\rm{cm}}^{-2}{\rm{s}}^{-1}$, suppressed by one order of magnitude compared to the benchmark value in Eq.~\eqref{eq:decay}. Thus, as we will show, our conservative upper bound on $\psi$ flux from DM decays already provides an useful insight into whether we could have a galactic flux of relativistic particles with keV energies comparable to the flux of solar axions.

\subsection{DM Annihilations}
\label{sec_ann}

Let us now consider the scenario where a fraction $f_\phi$ of DM can annihilate into relativistic $\psi$ particles. This includes two possible cases: DM directly annihilates into various particles including $\psi$, or it annihilates into some intermediate states, which subsequently decay to $\psi$. The goal here is similar to the study in the previous section. We want to derive some upper bounds on the total flux of $\psi$ on Earth from DM annihilations, assuming very general requirements for DM.

We assume that the abundance $\fp \Omega_\dm$ has already been set at a redshift $z_0$ via an unspecified mechanism about which we remain agnostic (e.g. freeze-out, freeze-in, or moduli decays). For redshifts below $z_0$ then, the average number density of $\phi$, $\np$, is given by:
\beq\label{nphi}
    \np(z) = \frac{\fp \Omega_\dm \rho_\crit}{\mp}(1+z)^3 \approx 1~\cm^{-3} \times \fp (1+z)^3 \bigg( \frac{\keV}{\mp} \bigg) \ , \quad z < z_0 \ ,
\eeq
where $\rho_\crit \approx 10^{-5} h^2$ GeV$\cdot$cm$^{-3}$ is the critical density of the Universe. We take $\Omega_\dm = 0.25$ and $h=0.7$\footnote{We are aware of the $H_0$ ``crisis'' in cosmology, so we split the difference.} as fiducial values for the cosmological parameters. We pick the benchmark scale of $\mp$ to be keV, motivated by the XENON1T excess.

Quite generically, the energy density in $\phi$ particles arises from a thermal bath (either the SM or a more exotic one, via either freeze-out or freeze-in). In order for this energy density to not be vanishingly small, the energy density of the bath itself must be non-negligible. Since after matter-radiation equality the energy density in all radiation, including this thermal bath, quickly dilutes, we are forced to have the $\phi$ relic abundance set before this happens. Thus, we arrive to the condition that
\beq\label{z0_condition}
    z_0 \gsim z_\eq \ ,
\eeq
where $z_\eq \approx 3000$ is the redshift of matter-radiation equality. Note that this is a very conservative condition. Cosmological data such as measurements of the acoustic peaks of the cosmic microwave background or of matter structure demand that the DM be very cold. For generic models this means that the redshift $z_0$ at which the DM abundance is fixed has to be much larger than $z_\eq$, before the smallest modes observed enter the horizon. However, we will restrict ourselves to the condition in Eq.~\eqref{z0_condition}.

A fixed abundance for $\phi$ implies an unchanging comoving number density. This results in the following consistency condition on the annihilation rate:
\beq\label{ann_rate}
	\Gamma_\mathrm{ann}(z_0) = \np(z_0) \sigv \lsim H(z_0) \ ,
\eeq
which guarantees the annihilations have stopped occurring. Taking the Hubble expansion rate as $H(z) = H_0 \sqrt{\Omega(z)} = 2 \times 10^{-18} ~ \sec^{-1} \sqrt{\Omega(z)}$, we arrive at
\bea\label{xsec_bound}
    \sigv & \lsim & 2 \times 10^{-18} ~\frac{\cm^3}{\sec} ~ \frac{\sqrt{\Omega(z_0)}}{(1+z_0)^3} \fp^{-1} \bigg( \frac{\mp}{\keV} \bigg) \nn\\
    & \approx & 10^{-18} ~\frac{\cm^3}{\sec} ~ \frac{1}{(1+z_0)^{3/2}} \fp^{-1} \bigg( \frac{\mp}{\keV} \bigg) \nn\\
    & \lsim & 7 \times 10^{-24} ~ \frac{\cm^3}{\sec}~ \fp^{-1} \bigg( \frac{\mp}{\keV} \bigg) \ ,
\eea
where in the second line, we approximate $\Omega(z_0) \approx \Omega(z=0) (1+z_0)^3$ in the matter-domination epoch, and for the last inequality we use Eq. \eqref{z0_condition}. We also want to emphasize that this is not a condition requiring $\phi$ to have a thermal relic abundance, but the annihilation of $\phi$ to become negligible after $z_{\eq}$, so that both the density background and perturbation evolutions are unaffected by DM annihilations into radiation (even when the radiation is not SM particles). Similar to the decay case, if DM annihilates into SM particles, there are usually much stronger constraints. Yet the bound could apply to cases in which a dark sector is completely secluded from the SM except for a portal through $\psi$, which is feebly coupled to the SM (e.g. axion).\footnote{For example, consider a dark sector containing $\phi$, a complex scalar DM candidate; $\chi_1, \tilde{\chi}_1$ and $\chi_2, \tilde{\chi}_2$, two vector-like pairs of fermions and $a$, an axion-like particle. We assume that all the particles are decoupled from the SM except for coupling to $a$, which couples to the SM electrons. The hidden sector particles are charged under a discrete $Z_3$ symmetry with the charge assignments as $\phi: +1, \;  \chi_1: +1, \;  \tilde{\chi}_1: -1, \; \chi_2: +1,  \; \tilde{\chi}_2: -1, a: 0$.
The Lagrangian (other than the kinetic terms) that respects the symmetry is
\beq
m_\phi^2 \phi^\dagger \phi + m_1 \chi_1 \tilde{\chi}_1 + m_2 \chi_2 \tilde{\chi}_2 + y_1 \phi \chi_1 \chi_1 + y_2 \phi \chi_2  \chi_2 + \frac{\partial^\mu a}{f_a} \chi_1^\dagger \bar{\sigma}_\mu \chi_2 +\frac{\partial^\mu a}{f_a} \tilde{\chi}_1^\dagger \bar{\sigma}_\mu \tilde{\chi}_2 + c.c.
\eeq
$\chi_1, \tilde{\chi}_1$ are degenerate in mass with mass $m_1$ and so are $\chi_2, \tilde{\chi}_2$ with a common mass $m_2$.
To have $\phi$ stable, we need to have the following mass inequality $m_2 < m_1 < m_\phi < 2 m_2$.
Then we have $\phi \phi^\dagger \to \chi_1 \chi_1^\dagger, \chi_1 \to a + \chi_2$ that leads to production of $a$'s.
}

We can repeat the same exercise for the case of a 3-body annihilation process. In this case, the consistency condition is given by:
\beq\label{ann_rate_3bdy}
    \Gamma_\mathrm{ann}(z_0) = \np^2(z_0) \langle \sigma v^2 \rangle_{3-\mathrm{body}} \lsim H(z_0) \ ,
\eeq
where $\langle \sigma v^2 \rangle_{3-\mathrm{body}}$ is the 3-body annihilation cross section. Then we have
\bea\label{3bdy_bound}
    \langle \sigma v^2 \rangle_{3-\mathrm{body}} & \lsim & 2 \times 10^{-18} ~ \frac{\cm^6}{\sec} ~ \frac{\sqrt{\Omega(z_0)}}{(1+z_0)^6} \fp^{-2} \bigg( \frac{\mp}{\keV} \bigg)^2 \nn\\
    & \lsim & 2 \times 10^{-34} ~ \frac{\cm^6}{\sec} ~ \fp^{-2} \bigg( \frac{\mp}{\keV} \bigg)^2 \ .
\eea

The flux of relativistic $\psi$ particles coming from the annihilating $\phi$ DM particles is thus given by:
\bea\label{flux2}
  \frac{d\Phi_2}{dE} & = & \left(\frac{dN_\psi}{dE}\right)_0 \frac{\fp^2 \sigv}{8 \pi \mp^2} ~ \int \rho_\dm^2(s)\, \di s \di \Omega \ , \qquad \qquad \; \; \; \text{2-body,} \\
   \frac{d\Phi_3}{dE} & = & \left(\frac{dN_\psi}{dE}\right)_0  \frac{\fp^3 \langle \sigma v^2 \rangle_{3-\mathrm{body}}}{24 \pi \mp^3} ~ \int \rho_\dm^3(s) \di s \di \Omega \ , \qquad \text{3-body.}
\eea
where, as in the case of decays, we evaluate the DM phase space integrals for each case separately, reporting them as $J_{\rm{ann}}^{2 \rightarrow 2}$ and $J_{\rm{ann}}^{3 \rightarrow 2}$ respectively in Appendix~\ref{app_Jfactor}. Considering the flux from the Milky Way and combining the equations above with the bounds from Eqs.~\eqref{xsec_bound} and \eqref{3bdy_bound}, we get:
\bea\label{flux2_bound}
    \Phi_2 & \lsim & 8 \times 10^{9} ~  \cm^{-2}\sec^{-1} ~ \fp \bigg( \frac{4~\keV}{\mp} \bigg) \left[\int \left(\frac{dN_\psi}{dE}\right)_0 dE\right],  \\
    \Phi_3 & \lsim & 7 \times 10^{5} ~ \cm^{-2}\sec^{-1}  ~ \fp \bigg( \frac{4~\keV}{\mp} \bigg) \left[\int \left(\frac{dN_\psi}{dE}\right)_0 dE\right]\ .\label{flux3_bound}
\eea

Finally, we remark on the interplay between $\fp$ and $\sigv$ which, under our assumptions so far, are independent of each other. One might attempt to evade the bounds above by allowing for a very large annihilation cross section $\sigv$ during matter-domination while at the same time jettisoning the requirement that the relic abundance of $\phi$ be set before matter-radiation equality. Effectively, this implies that $\Gamma_\mathrm{ann} > H$ during matter-domination and the comoving number density of $\phi$ decreases. In order to avoid cosmological constraints, one could then expect that the $\phi$ is only a subdominant component of all DM. However, since the $\phi$ particles are being constantly annihilated during matter-domination, the fraction $\fp$ of DM $\phi$ particles {\it today} will be tiny, proportional to $1/\sigv$, regardless of whether the annihilations ever stop (as in a standard freeze-out scenario) or not. Therefore, the flux $\Phi_2 \propto \fp^2 \sigv \propto 1/\sigv$ will be suppressed and this attempt ultimately fails.

\section{Implications for XENON1T}
\label{sec_xenon1T}

In this section, we will apply the bounds derived in the previous section to the possible explanation of the XENON1T excess via absorption of relativistic $\psi$ particles. As shown in Refs.~\cite{Aprile:2020tmw}, an explanation based on solar axions is in tension with stellar cooling bounds~\cite{PhysRevD98030001}. We will consider the scenario with $\psi$ from the galactic source, i.e., DM decay or annihilation of $\phi$ particles. Since the observed XENON1T excess lies around $\sim 2~\keV$, we then require $\mp \sim 4~\keV$. In this case $\psi$'s couplings to dark matter could be less constrained, yet we will show that even then the simple bounds we derive in the previous section rule out this scenario.

\subsection{Axions}
\label{sec_axions}
To use the solar axion, e.g., ABC axions, to explain the XENON1T excess, one needs to have axion-electron coupling $g_{ae} \sim $ (2--3)$\times 10^{-12}$~\cite{Aprile:2020tmw}, which is about one order of magnitude above the bound from cooling of red giant~\cite{Giannotti:2017hny}\footnote{For a handy compilation of the latest axion bounds, see \href{https://github.com/cajohare/AxionLimits}{github.com/cajohare/AxionLimits}.}. Now let us consider relativistic axions from DM decays or annihilations. In this case, the number of absorption events at direct detection is proportional to $\Phi_{\rm{DM}} \, g_{ae}^2$ with $\Phi_{\rm{DM}}$ being the flux of axions from DM decays or annihilations as contrasted to $g_{ae}^4$ for ABC axions (with two powers of $g_{ae}^2$ from solar production and two powers of $g_{ae}^2$ from absorption). Requiring $g_{ae}$ to satisfy the current red giant bound, we have the absorption rate at direct detection reduced by a factor of 100 compared to the ABC axion explanation. Thus we need $\Phi_{\rm{DM}} \sim 100 \Phi_{\rm{ABC}}(g_{ae} = 2\times 10^{-12})$ to be able to account for the excess.
For the ABC axion, the differential flux sharply peaks at (1--2) keV with~\cite{Redondo:2013wwa}
\beq
\left.\frac{d\Phi_\mathrm{ABC}}{dE}\right|_{\rm{peak}} \sim 4 \times 10^{11} \, {\rm{cm}^{-2}}{\rm{s}^{-1}}{\rm{keV}}^{-1}\left(\frac{g_{ae}}{2 \times10^{-12}}\right)^2.
\eeq
Integrating over $\frac{d\Phi_\mathrm{ABC}}{dE}$ over the relevant energy bins (1--7) keV, we find that a flux of
\beq
\Phi_{\rm{DM}} \sim 10^{14} \, {\rm{cm}}^{-2}{\rm{s}}^{-1}
\eeq
is required to explain the excess, a value at least three orders of magnitude above our estimated upper bounds in Eq.~\eqref{eq:decay} and Eqs.~\eqref{flux2_bound} and \eqref{flux3_bound}. Equivalently, this tells us that the maximal number of axion absorption events at XENON1T that could have a DM origin is $\sim 0.05$, assuming a value of $g_{ae}$ that saturates the red giant bound and a flux $\Phi_{\rm{DM}}$ saturating the bound in Eq.~\eqref{flux2_bound}.

\subsection{Dark Photon}
\label{sec_dphoton}
A dark photon could kinetically mix with ordinary photon through a coupling $\epsilon F^{\mu\nu}F^\prime_{\mu\nu}/2$~\cite{Holdom:1985ag}, where $\epsilon$ is the mixing parameter and $F, F^\prime$ are the field strengths of the $U(1)_{\rm{EM}}$ and $U(1)_d$ gauge groups respectively.

There are two possible scenarios of dark photons~\cite{An:2013yua}:
\begin{itemize}
\item Stueckelberg case (SC): this is the limit in which the dark Higgs responsible for the breaking of the dark $U(1)_d$ is so heavy that it is decoupled from the low energy effective field theory.
\item Higgs case (HC): in this scenario the dark Higgs is light and there is no decoupling.
\end{itemize}
They have quite different solar production and direct detection properties. On the one hand, in terms of production, we have~\cite{An:2013yua}:
\beq
{\rm {SC}}: \gamma^{(*)} \to \gamma^\prime; \quad {\rm {HC}}: \gamma^{(*)} \to \gamma^\prime h^\prime,
\eeq
where $\gamma^\prime$ is the dark photon and $h^\prime$ is the light dark Higgs. Note that for the HC scenario, it is an associated production and the Sun could produce both dark photons and dark Higgs. For the SC scenario, the production is dominated by resonances, which are effective in the $(10 - 300)$ eV energy range, after which bremsstrahlung dominates, yielding an exponentially decaying flux~\cite{An:2013yfc, Redondo:2013lna}. On the other hand, HC dark photons and dark Higgses have a pretty flat spectrum of flux extending to above keV~\cite{An:2013yua}.

On the other hand, in terms of direct detection, we have
\beq
{\rm {SC}}:  \gamma^\prime + {\rm{atom}} \to  {\rm{atom}}^++  e^-  ; \quad {\rm{HC}}:\gamma^\prime (h^\prime) + {\rm{atom}} \to h^\prime (\gamma^\prime) + {\rm{atom}}^+ e^-.
\eeq
Note that in the first case a dark photon is absorbed and ionizes the atom, while in the second case a dark photon (dark Higgs) is absorbed while a dark Higgs (dark photon) appears. In other words the HC is a scattering process. We will focus below on the SC scenario, which is the purely absorption case.

\begin{figure}[h!]
\centering
\includegraphics[width=0.7\textwidth]{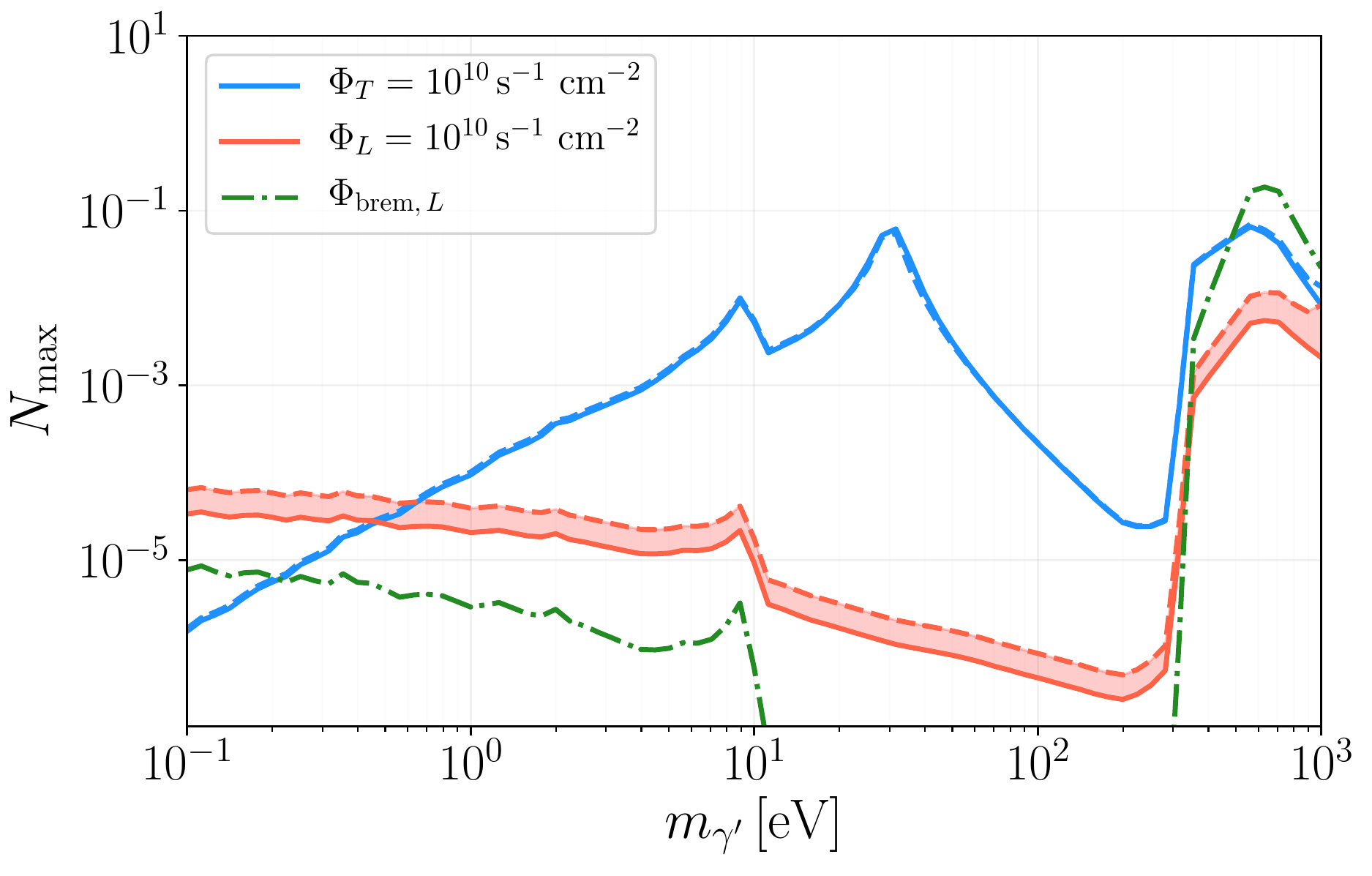}
\caption{The maximal number of events as a function of dark photon mass, with $\epsilon$ saturating its current upper bounds, for a total flux, for the transverse (blue) and longitudinal (red) modes, of $10^{10}~\sec^{-1}\cm^{-2}$. This corresponds to largest bound we found for the $\psi$ flux, Eq.~\ref{eq:decay}, taking $\int \left(\frac{dN_\psi}{dE}\right)_0 dE = 1$. The solid (dashed) lines represent the case of a monochromatic (box-shaped) spectrum centered at $E=2~\keV$ (between 1--4 keV). The dot-dashed green line corresponds to the bremsstrahlung flux of the longitudinal modes from the Sun.}
\label{fig:eventsmax}
\end{figure}

There are strong constraints on $\epsilon$ as a function of dark photon mass. The constraints for dark photon with mass around or below keV is nicely summarized in Fig.12 of ref.~\cite{Fabbrichesi:2020wbt}. The expected number of events, before accounting for the detector efficiency) as a function of incoming flux, $\epsilon$, and $m_{\gamma^\prime}$ has been computed in refs.~\cite{An:2013yua, An:2013yfc} (summarized in sec 3.2 of ref.~\cite{Bloch:2016sjj}). Assuming that $\epsilon$ saturates the current constraints for a given $m_{\gamma^\prime}$, we show the expected number of events at XENON1T (without taking considering detector efficiency) as a function of $m_{\gamma^\prime}$ in Fig.~\ref{fig:eventsmax}. We consider three possibilities for the flux of relativistic $\gamma^\prime$: \begin{itemize}
\item A monochromatic flux from DM decays or annihilations: $\frac{d \Phi_{\rm{DM}}}{dE} \propto \delta(E - E_0)$, where $E$ is the dark photon's energy, and we take $E_0 = 2$ keV, the energy where the excess is observed;
\item A box-shaped flux from  DM decays or annihilations: $\frac{d \Phi_{\rm{DM}}}{d E} \propto {\rm{rect}}(E)$, where the rectangular function satisfies ${\rm{rect}}(E) = 1$ when 1 keV $< E < 4$ keV and is zero everywhere else;
\item The solar flux from bremsstrahlung of dark photons in the Sun. The differential flux is given in Eq. (4.11) of ref.~\cite{Redondo:2013lna}. Note that it is exponentially suppressed when $E \gtrsim 0.3$ keV.
\end{itemize}
From Fig.~\ref{fig:eventsmax}, we see that the solar flux could only lead to at most 0.2 events. Even if we consider possible fluxes with DM origins, we could still have at most $\sim 0.1$ events assuming that the flux saturates the bounds in Eq.~\eqref{eq:decay} and Eqs.~\eqref{flux2_bound} and \eqref{flux3_bound}, regardless of whether the flux is monochromatic or box-shaped. We have taken $\int \left(\frac{dN_\psi}{dE}\right)_0 dE = 1$, which more generally could be of order $\sim \Ord(\text{few})$, without changing our conclusions.


\section{Caveats to our bounds}
\label{sec_caveats}

Although our upper bounds on flux of relativistic particles from DM decays or annihilations in Sec.~\ref{sec_bounds} are quite general, one could still devise complicated models to get around them. We will discuss one such possibility below.

For the DM annihilation scenario we implicitly assume that $\sigv$ is velocity-independent and does not change with time. Yet it is known that $\sigv$ could be velocity dependent, thereby changing as a function of redshift. In particular, if $\sigv = \sigv\vert_{z_0} (v/v_0)^{-n}$, where $v_0$ is the average velocity of the DM particles at $z_0$ (e.g. arising from adiabatic cooling in the case of a thermal origin), then the upper bound on the cross section given by Eq.~\eqref{xsec_bound} can be transformed into a bound on the cross section for DM particles in the galaxy (where $v\sim10^{-3}$):
\beq\label{sigv_galaxy}
    \sigv_\mathrm{MW}  &  \lsim & 10^{-18} ~ \frac{\cm^3}{\sec} ~\frac{(10^3 v_0)^n}{(1+z_0)^{3/2}} ~ \fp^{-1} \bigg( \frac{\mp}{\keV} \bigg) \ .
\eeq
We can see that for the $n=1$ case, or equivalently $\sigma \propto 1/v^2$, the upper bound on the cross section in the Milky Way, and therefore the flux of $\psi$ from DM annihilations, could be enhanced at most by a factor of $10^3$ (for $v_0 \sim 1$). Meanwhile, assuming the standard non-relativistic freeze-out value of $v_0\sim 0.1$ at $z_0 \approx 3000$, the enhanced flux becomes,  
\bea
\Phi_{\text{2-body},\, \sigma\propto 1/v^2} \, \approx \, 10^{12}\ \cm^{-2}\sec^{-1} ~ \fp \bigg( \frac{4~\keV}{\mp} \bigg) \left[\int \left(\frac{dN_\psi}{dE}\right)_0 dE\right],  \\
\Phi_{\text{2-body},\, \sigma\propto 1/v^4}\, \approx \, 10^{16}\ \cm^{-2}\sec^{-1} ~ \fp \bigg( \frac{4~\keV}{\mp} \bigg) \left[\int \left(\frac{dN_\psi}{dE}\right)_0 dE\right] \ . 
\eea 
Then, in principle, $1/v^n$-enhanced DM annihilations with $n > 1$ could produce a large enough new flux of relativistic bosons to account for the XENON1T excess. However, it's important to note that these annihilations will also be subject to additional observational constraints depending on the exact details of the particle physics model. For example, the $\sigma \propto 1/v^2$ behavior of the cross section is a characteristic feature of particle physics models with Sommerfeld enhancement and requires the presence of a light mediator. And as shown in ref.~\cite{Bringmann:2018jpr}, the enhanced cross section of such models during the Dark Ages is subject to various cosmological constraints. Thus, velocity enhanced annihilations serve as a promising production mechanism for relativistic DM particles, particularly in context of the XENON1T excess. We leave a detailed study of their cosmological and astrophysical implications to future work.

\section{Conclusion and Outlook}
\label{sec_disc}

Absorption of light bosons, DM or not, is an important target at direct detection. In this article, we explore a new scenario in which DM decays or annihilations could produce very weakly-coupled relativistic light bosons, axions or dark photons. We find several conservative upper bounds on the flux of the bosons at Earth as a function of DM mass and fraction, using simple requirements of DM that apply to generic models. These bounds are independent of the couplings and species of the bosons. With these bounds, one could show that adding this additional source of bosons with keV energy to the solar source could only lead to a small number of events ($\lesssim 1$) associated with absorption of bosons at direct detection, assuming that the bosons' coupling to the SM saturates current constraints. 

The simple bounds we derive will hold, regardless of whether the XENON1T excess survives scrutiny with data from upcoming experiments. For the present XENON1T excess, the bounds corner one class of possible explanations: absorption of relativistic bosons, assuming that all current constraints on the boson's coupling to standard model are correct. As usual, there could be loopholes to the bounds if one is willing to do more model building gymnastics, e.g., having a non-trivial time-dependent annihilation cross section; we defer that exercise to future work. 

Imposing the bounds we derive, the potential DM source of relativistic bosons could still be as important as or even dominate over the solar source. This raises the general question whether we could obtain some direction information of electron recoil events at direct detection to tell apart where the incoming particles come from. This could help narrow down possible explanations in case of a confirmed discovery in the future.

\paragraph{Acknowledgements}

We thank Savvas Koushiappas and Matt Reece for useful discussions and several critical comments. JF, MBA and JB are supported by the DOE grant DE-SC-0010010 and NASA grant 80NSSC18K1010.

\appendix

\section{Calculation of $J$ factors}
\label{app_Jfactor}
In this appendix, we detail the $J$-factor calculations and explain why we only consider the contributions to the flux from the MW halo.

For uniformity, in our calculations we assume a cuspy profile, the Navarro-Frenk-White (NFW) density profile~\cite{Navarro:1995iw}, for all DM regions including the MW halo and its dwarf galaxies,
\beq
\rho_{\rm{DM}}^{\rm {halo}}(r) = \frac{\rho_0}{\frac{r}{r_s} \left(1 + \frac{r}{r_s} \right)^2},
\eeq
where $r_s$ and $\rho_0$ are the scale radius and the mean DM density respectively. For the MW halo, $r_s = 20$ kpc and $\rho_0$ is determined such that $\rho(r_\odot) = 0.4$ GeV/cm$^3$~\cite{Sivertsson:2017rkp, Buch:2018qdr}. The $J$ factor integrals outlined in the main text can then be performed in a straightforward fashion over the entire solid angle for a chosen line-of-sight distance. 

In case of dwarfs, however, the story is slightly complicated by the fact that they are finite sized objects further away from the main halo. Following refs.~\cite{2009ApJ7041274W, Geringer-Sameth:2014yza, Evans:2016xwx}, the NFW parameters for a dwarf are $r_s = 5 R_h$ and $\rho_0 \sim M_h/r_s^3$, where $R_h$ denotes the half-light radius and $M_h = M(R_h)$ is the DM mass enclosed within it. To compute the $J$ factors, we also appropriately modify the phase space angle for an integration over a cylindrical volume,
\beq
d \Omega \, d s \rightarrow \frac{1}{D^2} \, 2 \pi R \, dR \, d z,
\eeq
where $D$ is the distance to the dwarf in the heliocentric frame, $z$ is the line-of-sight direction, and $R$ is a planar polar coordinate in the sky. Under these assumptions, as shown in ref.~\cite{Evans:2016xwx}, the $J$ factors can be computed analytically and have the following asymptotic form,
\beq
(J_{\rm{ann}}^{2\rightarrow2})_{\rm{dwarf}} \sim \frac{\rho_0^2 \, r_s^2 \, \theta_{\rm{max}}}{D}, \quad \quad  (J_{\rm{ann}}^{3\rightarrow2})_{\rm{dwarf}} \sim \frac{\rho_0^3 \, r_s^3}{D^2}, \quad \quad J_{\rm{dec}}^{\rm{dwarf}} \sim \frac{\rho_0 \, r_s^3}{D^2} \, \log \left( \frac{D \theta_{\rm{max}}}{2 r_s} \right),
\eeq
where $\theta_{\rm{max}}$ is the angle between the dwarf's centre and its outermost member star. Our results for the MW and several classical and ultra-faint dwarfs, adopting the values for all relevant parameters from ref.~\cite{Geringer-Sameth:2014yza}, are summarized in Table~\ref{table:d_j_factors}. For comparison, we also present results for the MW Galactic Center region, defined such that its comparable in volume to several dwarfs.
\begin{table}[!htb]
\centering
\begin{tabular}{ | c | c | c | c | c | c |}  \hline
DM region & \pbox{10cm}{ \vspace{2mm} Distance \\[2pt] \centering [kpc] \vspace{2mm}}  & \pbox{10cm}{ \vspace{2mm} $\rho_{0} $ \\[2pt] \centering [$M_\odot/{\rm {pc}}^3$] \vspace{2mm}} & \pbox{10cm}{ \vspace{2mm} $J_{\rm{dec}}$ \\[2pt] \centering [${\rm{GeV}}/{\rm {cm}}^2$] \vspace{2mm} } & \pbox{10cm}{ \vspace{2mm} $J_{\rm {ann}}^{2\rightarrow2}$  \\[2pt] \centering [${\rm {GeV}}^2/{\rm{cm}}^5$] \vspace{2mm} } & \pbox{10cm}{ \vspace{2mm} $J_{\rm{ann}}^{3\rightarrow2}$  \\[2pt] \centering [${\rm {GeV}}^3/{\rm{cm}}^8$] \vspace{2mm} }  \\ \hline
MW halo (full) & -- &  -- & $2.1 \times 10^{23}$ & $1.2 \times 10^{23}$ & $ 1.0 \times 10^{24}$ \\[2pt]
MW Galactic Center & $\sim 8$ & $\sim 11$ & $\sim 7.8 \times 10^{21}$ & $\sim 1.6 \times 10^{22}$ & $\sim 4.4 \times 10^{23}$ \\[2pt]
Carina & 105 & $0.10$ & $1.1 \times 10^{17}$ & $5.6 \times 10^{18}$  & $5.2 \times 10^{17}$ \\[2pt]
Draco & 76 & $0.30$ & $3.1 \times 10^{17}$ & $3.2 \times 10^{19}$  & $ 7. \times 10^{18}$ \\[2pt]
Fornax & 147 & $0.04$ & $3.9 \times 10^{17}$ &$9.7 \times 10^{18}$ & $ 3.2 \times 10^{17}$ \\[2pt]
Ursa Minor & 76 & $0.16$ &  $5.3 \times 10^{17}$ & $1.9 \times 10^{20}$ & $ 1.0 \times 10^{20}$\\[2pt]
Segue I & 23 & $3.03$ &  $1.5 \times 10^{17}$ & $1.0 \times 10^{20}$ & $ 5.8 \times 10^{20}$ \\[2pt]
Ursa Major II & 32 & $0.31$ &  $5.6 \times 10^{17}$ & $2.1 \times 10^{19}$ & $ 2.5 \times 10^{19}$ \\[2pt]
Willman I & 28 & $4.12$ & $5.45 \times 10^{16}$  & $1.2 \times 10^{20}$ & $ 3.4 \times 10^{20}$ \\[2pt] \hline
\end{tabular}
\caption{$J_{\rm {ann}}^{2\rightarrow2}$, $J_{\rm {ann}}^{3\rightarrow2}$ and $J_{\rm{dec}}$ factors for the MW halo and its dwarf galaxies. For properties of dwarfs, including distances and mean DM densities for a NFW profile, the parameters are adopted from ref.~\cite{Geringer-Sameth:2014yza}. The MW Galactic Center is defined as a galactocentric region with radius $100$ pc.}
\label{table:d_j_factors}
\end{table}

From the discussion above and Table~\ref{table:d_j_factors}, we conclude that:
\begin{itemize}
\item contribution of the MW halo is the dominant component of the daughter particle flux from DM annihilations and decays. Note that the full halo contribution is larger than the that of the Galactic Center, indicating that the $J$ factors do not strongly depend on the choice of our dark matter profile.
\item $J_{\rm {ann}}$ and $J_{\rm{dec}}$ factors for the MW Galactic Center are at least two orders of magnitude greater than those for a nearby, DM dense dwarf with a similar volume like Ursa Major II, both due to its proximity to us and greater mean DM density. We also note that the list of dwarfs presented here is illustrative, and questions regarding the DM density profile of several candidates, Segue I and Willman I in particular, are yet to be settled in the literature~\cite{2016MNRAS.462..223B}.
\end{itemize}

\bibliography{references_xe1t}
\bibliographystyle{utphys}

\end{document}